\documentclass[twocolumn,aps,showpacs,amsmath,superscriptaddress]{revtex4-1}

\usepackage{bm}
\usepackage{amsmath}
\usepackage{amssymb}
\usepackage[usenames,dvipsnames]{color}
\usepackage{graphicx}
\usepackage{dcolumn}
\usepackage{simplewick}
\usepackage{hyperref}
\usepackage{natbib}
\hypersetup{
 colorlinks=false,
 pdfborder={0 0 0}
 }

\begin{document}

\title{Electric dipole polarizability from perturbed Relativistic 
       Coupled-Cluster Theory: application to Neon}
\author{S. Chattopadhyay}
\affiliation{Physical Research Laboratory,
             Ahmedabad 380009, Gujarat,
             India}
\author{B. K. Mani}
\affiliation{Department of Physics, University of South Florida, Tampa,
             Florida 33620, USA}
\author{D. Angom}
\affiliation{Physical Research Laboratory,
             Ahmedabad 380009, Gujarat,
             India}
\begin{abstract}

  We develop a method based on the relativistic coupled-cluster theory to 
incorporate a perturbative interaction to the no-pair Dirac-Coulomb atomic 
Hamiltonian. The method is general and suitable to incorporate any 
perturbation Hamiltonian in a many electron atom or ion. Using this perturbed 
relativistic coupled-cluster (PRCC) theory we calculate the electric dipole 
polarizability, $\alpha$, of Neon. The linearized PRCC results are in very 
good agreement with the experimental value. However, the results of the
nonlinear PRCC shows larger uncertainty but it is consistent with the
observations from earlier works.
\end{abstract}

\pacs{31.15.bw,31.15.ap,31.15.A-,31.15.ve}


\maketitle


\section{Introduction}
The electric dipole polarizability, $\alpha$,  of an atom or molecule is an 
important property as it describes the interaction with an external electric 
field. Knowing accurate value of atomic $\alpha$ is imperative in different 
areas of research. To mention a few,  the parity nonconservation in atoms 
\cite{khriplovich-91} and the ultracold atoms 
\cite{anderson-95, bradley-95, davis-95} are of current interest. In short, 
knowledge of $\alpha$ is essential in any experiment which involves the atom 
interacting with an external electric field. Theoretically, $\alpha$ is 
calculated using the first order time-independent perturbation theory. A 
recent review \cite{mitroy-09} gives a detailed account of the current status 
of the theoretical results of atomic and ionic polarizabilities. An excellent
source of information related to atomic polarizability is ref.
\cite{maroulis-06}.  

  In the present work we extend the standard relativistic coupled-cluster
(RCC) theory to include an additional perturbation Hamiltonian. Here, it must 
emphasized that the coupled-cluster (CC) theory \cite{coester-58, coester-60} 
is one of the most reliable quantum many body theory. We refer to the CC 
theory with the additional perturbation as PRCC method. The theory is 
flexible and formulated to incorporate multiple perturbations which may have 
different tensor structures in the electronic sector. The perturbation
could be internal, like hyperfine interaction or external like the 
static electric field. For the present work, as $\alpha$ is the quantity of
interest, we take an external electric field as the perturbation. The 
coupled-cluster theory has been widely used for atomic\cite{nataraj-08,pal-07},
molecular\cite{isaev-04}, nuclear \cite{hagen-08}, condensed matter physics
\cite{bishop-09} calculations. The previous approaches of CC
theory for atomic calculations are conceptually different from the PRCC 
method in one aspect. In PRCC, the CC single and double (CCSD) excitation 
operators within the electron sector could be tensor operators of higher ranks. 
In which case the associated angular factors are complicated and are
evaluated using diagrammatic method \cite{lindgren-86}.  The PRCC equations
are, at first glance, complicated in appearance as it involves two sets of
cluster operators. However, the equations are linear and not nonlinear as in
the full RCC theory. For calculations, we used the equivalent of standard
CC single and double but the theory can be extended to higher orders.  

  The paper is organized as follows. In the Section. II we provide a brief 
description of the PRCC method and examine the structure of the cluster 
equations. This is followed with a short presentation of the details of
the numerical aspects and actual models used in Section. III. The Section. IV
discusses the results and we end with the conclusions.


\section{Theoretical methods}

The Dirac-Coulomb Hamiltonian $H^{\rm DC} $ , consisting of relativistic 
one-electron terms and electrostatic electron-electron interaction, is the 
starting point of our calculation. For an $N$-electron atom 
\begin{equation}
   H^{\rm{DC}} = \sum_{i=1}^N[c\bm{\alpha}_i \cdot \mathbf{p}_i + 
                 (\beta_i -1)c^2 - V_{N}(r_i)] + \sum_{i<j}\frac{1}{r_{ij}},
\end{equation}
where, $\bm{\alpha}_i $'s and the $\beta_i$ are the Dirac matrices, $V_N(r_i)$
is the nuclear potential and the last term represents the electron-electron 
Coulomb interactions. For a closed-shell atom the eigen-value equation is
\begin{equation}
   H^{\rm{DC}}|\Psi_{i}\rangle = E_{i}|\Psi_{i}\rangle .
\end{equation}
In the RCC theory the ground state wave-function of an atom is written as
\begin{equation}
   |\Psi_o\rangle = e^{T^{(0)}}|\Phi_o\rangle ,  
\end{equation} 
here $|\Phi_o\rangle$ is the ground state reference state. In the
CCSD approximation the cluster operator $T^{(0)} = T_1^{(0)} + T_2^{(0)}$. 
The advantage of CC theory is, it is an all order non-perturbative theory and
ensures that electron-electron correlation diagrams is accounted fully.
For future reference $T^{(0)}$ are referred to as the  unperturbed cluster 
operator. In the CCSD approximation the amplitude equations are written as
\begin{eqnarray}
   \langle\Phi_a^p|\bar{H}_{\rm{N}}|\Phi_0\rangle       & = & 0, \\
   \langle\Phi_{ab}^{pq}|\bar{H}_{\rm{N}}|\Phi_0\rangle & = & 0.
\end{eqnarray} 
Here $\bar{H}_{\rm{N}} = e^{-T^{(0)}}H_{\rm{N}}e^{T^{(0)}}$ is the similarity 
transformed Hamiltonian and 
$H_{\rm{N}}= H^{\rm{DC}} - \langle\Phi_0|H^{\rm{DC}}|\Phi_0\rangle$ is the 
normal ordered Hamiltonian\cite{bartlett-07}.

  In the presence of an additional interaction, which could be an internal
or external perturbation Hamiltonian $H' $, $|\Psi_o\rangle $ is modified. 
Examples of internal perturbations are hyper-fine interaction, Breit 
interaction, etc. And examples of external perturbations are Stark effect, 
Zeeman effect, etc. The perturbation Hamiltonian, in the electron sector, can 
be a tensor operator of rank one or higher. The effect of these perturbations, 
although important, are less in magnitude than the residual Coulomb 
interaction and a first order treatment is sufficient. To incorporate  $H'$,
we have developed a perturbed RCC (PRCC) theory.

   For the present work we take an external electric field as the perturbation
in an N-electron atom. The atomic Hamiltonian is then modified to
\begin{equation}
   H = H^{\rm{DC}} + \lambda H_{\rm{int}},
\end{equation}
where $ H_{\rm{int}} = \sum_{i=1}^{N} \mathbf{r}_{i}.\mathbf{E}$, with 
$\mathbf{E} $ as the external electric field and $\lambda$ is the perturbation
parameter. Eigen-value equation with the modified Hamiltonian is
$ H|\tilde\Psi_{0}\rangle = \tilde{E}_i |\tilde\Psi_{0}\rangle$.
In the PRCC theory the new ground state wave-function is 
\begin{equation}
   |\tilde{\Psi}_{0}\rangle = e^{T^{(0)}+\lambda \mathbf{T}^{(1)}}|\Psi_{0}
   \rangle.
\end{equation}
Here we have introduced the PRCC operator $\mathbf{T}^{(1)}$, 
which incorporates the many-body effects of $H_{\rm int}$. It operates on the
electronic sector and is a rank one operator as the electronic part of 
$H_{\rm int }$ is a tensor operator of rank one. To first order in
$\lambda$ we have 
$|\tilde{\Psi}_{0}\rangle = e^{T^{(0)}}[1 + \lambda \mathbf{T}^{(1)}]|\Psi_{0}
\rangle$ and in the CCSD approximation
$\mathbf{T}^{(1)} = \mathbf{T}_1^{(1)} + \mathbf{T}_2^{(1)}$.
More detailed description on the structures of  $\mathbf{T}_1^{(1)}$ and
$\mathbf{T}_2^{(1)}$ operators are given our previous publication 
\cite{mani-11-3}.

The PRCC amplitude are solutions of the equations 
\begin{eqnarray}
  && \langle \Phi_a^p |\left [ \contraction[0.5ex]{}{H}{_{\rm N}}{T}H_{\rm N}
     \mathbf{T}^{(1)} + \contraction[0.5ex]{}{H}{_{\rm N}}{T} 
     \contraction[0.8ex]{}{H} {_{\rm N}T^{(0)}}{T} H_{\rm N}T^{(0)}
     \mathbf{T}^{(1)} + \frac{1}{2!} \contraction[0.5ex]{}{H}{_{\rm N}}{T}
     \contraction[0.8ex]{}{H}{_{\rm N}T^{(0)}}{T} 
     \contraction[1.1ex]{}{H}{_{\rm N}T^{(0)}T^{(0)}}{T} 
     H_{\rm N}T^{(0)}T^{(0)}\mathbf{T}^{(1)} \right .
                              \nonumber \\
  && \left . + \frac{1}{3!} \contraction[0.5ex]{}{H}{_{\rm N}}{T}
     \contraction[0.8ex]{}{H}{_{\rm N}T^{(0)}}{T} 
     \contraction[1.1ex]{}{H}{_{\rm N}T^{(0)}T^{(0)}}{T} 
     \contraction[1.4ex]{}{H}{_{\rm N}T^{(0)}T^{(0)}T^{(0)}}{T} 
     H_{\rm N}T^{(0)}T^{(0)}T^{(0)}\mathbf{T}^{(1)} 
     \right ] |\Phi_0\rangle 
     =  - \langle\Phi_a^p | \left [
     \contraction[0.5ex]{}{H}{_{\rm int}}{T}H_{\rm int}T^{(0)} \right .
                     \nonumber \\
  && \left .   + \frac{1}{2!} \contraction[0.5ex]{}{H}{_{\rm int}}{T}
     \contraction[0.8ex]{}{H}{_{\rm int}T^{(0)}}{T} H_{\rm int}T^{(0)}T^{(0)}
     \right ]  |\Phi_0\rangle ,  
           \label{prcc_eq1}     \\
  && \langle \Phi_{ab}^{pq} |\left [ \contraction[0.5ex]{}{H}{_{\rm N}}
     {T}H_{\rm N}\mathbf{T}^{(1)} + \contraction[0.5ex]{}{H}{_{\rm N}}{T} 
     \contraction[0.8ex]{}{H} {_{\rm N}T^{(0)}}{T} H_{\rm N}T^{(0)}
     \mathbf{T}^{(1)} + \frac{1}{2!} \contraction[0.5ex]{}{H}{_{\rm N}}{T}
     \contraction[0.8ex]{}{H}{_{\rm N}T^{(0)}}{T} 
     \contraction[1.1ex]{}{H}{_{\rm N}T^{(0)}T^{(0)}}{T} 
     H_{\rm N}T^{(0)}T^{(0)}\mathbf{T}^{(1)}  \right .
                     \nonumber \\
  && \left . + \frac{1}{3!}    
     \contraction[0.5ex]{}{H}{_{\rm N}}{T}
     \contraction[0.8ex]{}{H}{_{\rm N}T^{(0)}}{T} 
     \contraction[1.1ex]{}{H}{_{\rm N}T^{(0)}T^{(0)}}{T} 
     \contraction[1.4ex]{}{H}{_{\rm N}T^{(0)}T^{(0)}T^{(0)}}{T} 
     H_{\rm N}T^{(0)}T^{(0)}T^{(0)}\mathbf{T}^{(1)} 
    \right ] |\Phi_0\rangle  =  - \langle\Phi_{ab}^{pq} | \left [
     \contraction[0.5ex]{}{H}{_{\rm int}}{T}H_{\rm int}T^{(0)} \right .
                      \nonumber \\
  && \left .  + \frac{1}{2!} \contraction[0.5ex]{}{H}{_{\rm int}}{T}
     \contraction[0.8ex]{}{H}{_{\rm int}T^{(0)}}{T} H_{\rm int}T^{(0)}T^{(0)}
     \right ]  |\Phi_0\rangle ,
           \label{prcc_eq2}   
\end{eqnarray}
where, $\contraction[0.5ex]{}{A}{}{B}AB $ represents all possible contractions
between the two operators $A$ and $B$. These are non-linear coupled equations. 
However, $T^{(0)}$ are solutions of the unperturbed RCC equations 
\cite{mani-09}, which are solved separately. With this consideration, the 
$\mathbf{T}^{(1)}$ equations are reduce to a set of linear algebraic equations.
The terms on the left hand side of Eq. (\ref{prcc_eq1}) and (\ref{prcc_eq2}) 
can be simplified to
\begin{eqnarray}
&&\{\contraction[0.5ex]{}{H}{_{\rm N}}{T}H_{\rm N}\mathbf{T}^{(1)}\}  = 
\{\contraction[0.5ex]{}{H}{_{\rm N}}{T}H_{\rm N}\mathbf{T}_1^{(1)}\} +
\{\contraction[0.5ex]{}{H}{_{\rm N}}{T}H_{\rm N}\mathbf{T}_2^{(1)}\},
                    \nonumber  \\
&&\{\contraction[0.5ex]{}{H}{_{\rm N}}{T} 
  \contraction[0.8ex]{}{H}{_{\rm N}T^{(0)}}{T} H_{\rm N}T^{(0)}
  \mathbf{T}^{(1)}\} = \{\contraction[0.5ex]{}{H}{_{\rm N}}{T}
  \contraction[0.8ex]{}{H}{_{\rm N}T^{(0)}}{T} 
   H_{\rm N}T_1^{(0)}\mathbf{T}_1^{(1)}\} 
+ \{\contraction[0.5ex]{}{H}{_{\rm N}}{T}
  \contraction[0.8ex]{}{H}{_{\rm N}T^{(0)}}{T} 
   H_{\rm N}T_1^{(0)}\mathbf{T}_2^{(1)}\} 
                    \nonumber \\
&& \;\;\;\;\;\;\;\;\;\;
 +\{\contraction[0.5ex]{}{H}{_{\rm N}}{T}
  \contraction[0.8ex]{}{H}{_{\rm N}T^{(0)}}{T} 
  H_{\rm N}T_2^{(0)}\mathbf{T}_1^{(1)}\} +
\{\contraction[0.5ex]{}{H}{_{\rm N}}{T}
  \contraction[0.8ex]{}{H}{_{\rm N}T^{(0)}}{T} 
  H_{\rm N}T_2^{(0)}\mathbf{T}_2^{(1)}\},
                   \nonumber     \\
&&\{\contraction[0.5ex]{}{H}{_{\rm N}}{T}
  \contraction[0.8ex]{}{H}{_{\rm N}T^{(0)}}{T} 
  \contraction[1.1ex]{}{H}{_{\rm N}T^{(0)}T^{(0)}}{T} 
  H_{\rm N}T^{(0)}T^{(0)}\mathbf{T}^{(1)}\} =  
\{\contraction[0.5ex]{}{H}{_{\rm N}}{T}
  \contraction[0.8ex]{}{H}{_{\rm N}T^{(0)}}{T} 
  \contraction[1.1ex]{}{H}{_{\rm N}T^{(0)}T^{(0)}}{T} 
  H_{\rm N}T_1^{(0)}T_1^{(0)}\mathbf{T}_1^{(1)}\}
                   \nonumber    \\
&& \;\;\;\;\;\;\;\;\;\;
   + \{\contraction[0.5ex]{}{H}{_{\rm N}}{T}
  \contraction[0.8ex]{}{H}{_{\rm N}T^{(0)}}{T} 
  \contraction[1.1ex]{}{H}{_{\rm N}T^{(0)}T^{(0)}}{T} 
  H_{\rm N}T_1^{(0)}T_2^{(0)}\mathbf{T}_1^{(1)}\}
+ \{\contraction[0.5ex]{}{H}{_{\rm N}}{T}
  \contraction[0.8ex]{}{H}{_{\rm N}T^{(0)}}{T} 
  \contraction[1.1ex]{}{H}{_{\rm N}T^{(0)}T^{(0)}}{T} 
  H_{\rm N}T_1^{(0)}T_1^{(0)}\mathbf{T}_2^{(1)}\}, 
                 \nonumber 
\end{eqnarray}
where, $\{ A \} $ represents the normal order form of the operator $A$.
Similarly, the terms on the right hand side expand to
\begin{eqnarray}
\{\contraction[0.5ex]{}{H}{_{\rm int}}{T}H_{\rm int}T^{(0)} \} & = &
\{\contraction[0.5ex]{}{H}{_{\rm int}}{T}H_{\rm int}T_1^{(0)} \} + 
\{\contraction[0.5ex]{}{H}{_{\rm int}}{T}H_{\rm int}T_2^{(0)} \}, 
               \nonumber    \\
\{\contraction[0.5ex]{}{H}{_{\rm int}}{T}
 \contraction[0.8ex]{}{H}{_{\rm int}T^{(0)}}{T} H_{\rm int}T^{(0)}T^{(0)}\} &=&
\{\contraction[0.5ex]{}{H}{_{\rm int}}{T}
 \contraction[0.8ex]{}{H}{_{\rm int}T^{(0)}}{T} H_{\rm int}T_1^{(0)}T_1^{(0)}\}
  + \{\contraction[0.5ex]{}{H}{_{\rm int}}{T}
 \contraction[0.8ex]{}{H}{_{\rm int}T^{(0)}}{T} H_{\rm int}T_1^{(0)}T_2^{(0)}\}.
\nonumber
\end{eqnarray}
For further evaluation we use diagrammatic techniques to calculate the 
contributions from these terms. In total there are 42 single diagrams
and 102 doubles diagrams. 

   From the time independent perturbation theory the dipole polarizability of 
an atom is 
\begin{equation}
  \alpha = -2  \sum_{I} \frac
  {\langle \Psi_0|\mathbf D|\Psi_I\rangle \langle \Psi_I|\mathbf D|
  \Psi_0\rangle}{E_0 - E_I},
\end{equation}
where, $|\Psi_I\rangle$ are the intermediate atomic states. In the PRCC theory 
the sum over states is implicit and $\alpha$ is the expectation value of the 
operator $e^{{\mathbf{T}^{(1)}}^\dagger}De^{T^{(0)}} 
+ e^{{T^{(0)}}^\dagger}De^{\mathbf{T}^{(1)}}$. After expansion, collecting 
terms upto second order in cluster operators
\begin{eqnarray}
   \alpha & = & \frac{1}{N}\langle \Phi_0 |
     \{\contraction[0.5ex]{}{T}{_{1}^{(1)\dagger}}{D}
     \mathbf{T}_1^{(1)\dagger}\mathbf{D} \}
     + \{\contraction[0.5ex]{}{D}{T}{}\mathbf{D}\mathbf{T}_1^{(1)}\} 
     + \{ \contraction[0.5ex]{}{\mathbf{T}}{_1{^{(1)\dagger}}}{D}
     \contraction[0.8ex]{}{\mathbf{T}}{{_1{^{(1)\dagger}}}D}{T}
     \mathbf{T}_1{^{(1)\dagger}}\mathbf{D}T_2^{(0)} \} 
             \nonumber \\
     && 
     + \{ \contraction[0.5ex]{}{T}{_2{^{(0)\dagger}}}{D}
     \contraction[0.8ex]{}{T}{{_2{^{(0)\dagger}}}D}{T}
     {T}_2{^{(0)\dagger}}\mathbf{D}\mathbf{T}_1^{(1)} \}
     + \{ \contraction[0.5ex]{}{\mathbf{T}}{_1{^{(1)\dagger}}}{D}
     \contraction[0.8ex]{}{\mathbf{T}}{{_1{^{(1)\dagger}}}D}{T}
     \mathbf{T}_1{^{(1)\dagger}}\mathbf{D}T_1^{(0)} \}
     + \{ \contraction[0.5ex]{}{T}{_1{^{(0)\dagger}}}{D}
     \contraction[0.8ex]{}{T}{{_1{^{(0)\dagger}}}D}{T}
     {T}_1{^{(0)\dagger}}\mathbf{D}\mathbf{T}_1^{(1)} \} 
             \nonumber \\
     && 
     +\{ \contraction[0.5ex]{}{\mathbf{T}}{_2{^{(1)\dagger}}}{D}
     \contraction[0.8ex]{}{\mathbf{T}}{{_2{^{(1)\dagger}}}D}{T}
     \mathbf{T}_2{^{(1)\dagger}}\mathbf{D}T_1^{(0)} \}
     +\{ \contraction[0.5ex]{}{T}{_1{^{(0)\dagger}}}{D}
     \contraction[0.8ex]{}{T}{{_1{^{(0)\dagger}}}D}{T}
     {T}_1{^{(0)\dagger}}\mathbf{D}\mathbf{T}_2^{(1)} \}
     + \{ \contraction[0.5ex]{}{\mathbf{T}}{_2{^{(1)\dagger}}}{D}
     \contraction[0.8ex]{}{\mathbf{T}}{{_2{^{(1)\dagger}}}D}{T}
     \mathbf{T}_2{^{(1)\dagger}}\mathbf{D}T_2^{(0)} \} 
             \nonumber \\ 
     && 
     + \{ \contraction[0.5ex]{}{T}{_2{^{(0)\dagger}}}{D}
     \contraction[0.8ex]{}{T}{{_2{^{(0)\dagger}}}D}{T}
     {T}_2{^{(0)\dagger}}\mathbf{D}\mathbf{T}_2^{(1)} \}
     | \Phi_0\rangle , 
   \label{alpha_eqn}
\end{eqnarray}
where $N = \langle \Phi_0| \{ \contraction[0.5ex]{}{T}{_1{^{(0)\dagger}}}{T}
           T_1{^{(0)\dagger}}T_1{^{(0)}}\} 
           +\{ \contraction[0.5ex]{}{T}{_2{^{(0)\dagger}}}{T}
           T_2{^{(0)\dagger}}T_2{^{(0)}}\}|\Phi_0\rangle $
is the normalization constant. The unperturbed wave-function is normalized, 
but we have to normalize the perturbed wave-function.


\section{Calculational Details}

  Using PRCC we have done a systematic study of dipole polarizability of Neon. 
For precision atomic theory calculation the orbital basis set is an important 
factor. In our calculations we use even tempered Gaussian type orbitals (GTOs)
\cite{mohanty-90}. The radial part of the Dirac bi-spinor is  expressed in 
terms of the linear combination of GTOs. The GTOs  
\begin{equation}
   g_{\kappa p}^{L}(r) = C^{L}_{\kappa i} r^{n_{\kappa}}e^{-\alpha_{p}r^{2}}.
\end{equation}
The index $p$ is the number of basis functions. The exponent $\alpha_{p}$ 
depends on two parameters $\alpha_{0}$ and $\beta$ and are related as
$   \alpha_{p} = \alpha_{0} \beta^{p-1}$, where $p=0,1\ldots m$ and $m$ is the 
number of $ g_{\kappa p}^{L}(r)$ considered. For present calculations we use
even tempered basis, where  $\alpha_0 $ and $\beta$ are unique for each of
the symmetries. The symmetry wise values of the parameters are listed in 
Table. \ref{ne_basis}. 
\begin{table}[h]
	\caption{The GTO even tempered basis parameters used in the present
                 calculations.}
\label{ne_basis}
   \begin{tabular}{lcccccc}
      \hline
   Symmetry   & $s_{1/2}$ & $p_{1/2}$ & $p_{3/2}$ & $d_{3/2}$ & $d_{5/2}$ 
              & $f_{5/2}$  \\
   $\alpha_{0}$ & 0.0925    & 0.1951    & 0.1917    & 0.0070    & 0.0070 
              & 0.0069     \\
   $\beta$    & 1.45      & 2.71      & 2.71      & 2.70      & 2.70  
              & 2.69        \\
   Symmetry   & $f_{7/2}$ & $g_{7/2}$ & $g_{9/2}$  & & & \\
   $\alpha_{0}$ & 0.0069    & 0.0069    & 0.0069     & & & \\
   $\beta$    & 2.69      & 2.69      & 2.69       & & & \\
            \hline
 \end{tabular}
\end{table}
Although, in principle, a complete set of orbitals are needed, it is 
near impossible to go beyond a few hundred. Even at a few hundred the 
computational requirements is very high. Another practical consideration,
with further increase the gain in accuracy is marginal or non-existent once
the basis set converges. The basis parameters are optimized such that the 
core orbital energies are in good agreement with the GRASP92 \cite{parpia-96}
results. For information, the orbital energies are listed in 
Table. \ref{ne_energy}.
\begin{table}[h]
    \caption{Orbital energies of obtained from GRASP92 and GTO in 
             atomic units.}
   \label{ne_energy}
   \begin{tabular}{ldddd}
   \hline
     Orbital  & \multicolumn{1}{r}{$1s_{1/2}$} & 
                \multicolumn{1}{r}{$2s_{1/2}$} & 
                \multicolumn{1}{r}{$2p_{1/2}$} & 
                \multicolumn{1}{r}{$2p_{3/2}$}  \\ \hline
     GTO      & -32.8177   &  -1.9357   & -0.8526    & -0.8480     \\
     GRASP92  & -32.8145   &  -1.9387   & -0.8528    & -0.8482     \\
             \hline
   \end{tabular}
\end{table}

  To test and check the theory, we consider the linearized PRCC theory. There 
are then 10 singles and 10 doubles diagrams each. However, only 6 of the 
singles diagram but all the doubles diagrams contribute when DHF orbitals are 
used. Detailed descriptions of the diagrammatic calculations are given
in ref. {\cite{mani-11-3}}. The result, along with previous and experimental
values, are given in Table. \ref{ne_result1}. It shows that our results agrees 
very well with the experimental data and  indicates that the PRCC theory, even
at the linear level, gives  accurate results for a single reference system 
like Ne.
\begin{table}[h]
  \caption{Contribution from Linearized PRCC to the static dipole 
           polarizability of Ne  and comparison with previous results.}
  \label{ne_result1}
  \begin{tabular}{lcccc} \hline
     This work & CCSDT\cite{soldan-01} & RCCSDT\cite{nakajima-01} & 
      MBPT\cite{thakkar-92} & Expt.\cite{holm-90}  \\ \hline
      2.6695 & 2.6648 & 2.697 & 2.665 & 2.670(5)  \\ \hline
  \end{tabular}
\end{table}
To analyze the impact of basis set truncation, we examine
the convergence of $\alpha $ with the size of basis set. For this we start
with a basis set of 50 GTOs and do a series of calculations by increasing the
basis size in steps. The value of $ \alpha$ converges to 2.6695 when the basis 
set size is 124. However, for confirmation we increase the basis set size 
upto 171 and results are listed in Table. \ref{ne_result2}.
\begin{table}[h]
  \caption{Convergence pattern of $\alpha$ of Ne as a function of 
           the Basis set size.}
  \label{ne_result2}
  \begin{center}
  \begin{tabular}{lcc}
      \hline
      No. of orbitals & Basis size & Polarizability  \\
      \hline
      50 &  $(10s, 6p,  6d,   4f,   4g)$ & 2.7279  \\
      60 &  $(12s, 7p,  7d,   5f,   5g)$ & 2.7087  \\
      75 &  $(13s, 9p,  9d,   7f,   6g)$ & 2.6849  \\
      91 &  $(15s, 11p, 11d,  8f,   8g)$ & 2.6712  \\
      108 & $(20s, 13p, 11d,  11f,  9g)$ & 2.6696  \\
      124 & $(22s, 14p, 14d,  13f, 10g)$ & 2.6695  \\
      145 & $(27s, 17p, 16d,  14f, 12g)$ & 2.6695  \\
      163 & $(29s, 21p, 17d,  16f, 13g)$ & 2.6695  \\
      171 & $(31s, 23p, 18d,  16f, 13g)$ & 2.6695  \\
      \hline
  \end{tabular}
  \end{center}
\end{table}
In this calculation we have considered finite size Fermi density distribution
for the nucleus.
\begin{equation}
   \rho_{\rm nuc}(r) = \frac{\rho_0}{1 + e^{(r-c)/a}}.
\end{equation}
Here, $a=t 4\ln(3)$. The parameter $c$ is the half charge radius so that 
$\rho_{\rm nuc}(c) = {\rho_0}/{2}$ and $t$ is the skin thickness. The PRCC 
equations are solved iteratively using Jacobi method, we have chosen this 
method as it is parallelizable. The method, however, is slow to converge. So, 
we use direct inversion in the iterated subspace (DIIS)\cite{pulay-80} to 
accelerate the convergence.


\section{Results and discussions}

 In the properties calculations the CC expression of the polarizability 
operator, $e^{{\mathbf{T}^{(1)}}^\dagger}De^{T^{(0)}} 
+ e^{{T^{(0)}}^\dagger}De^{\mathbf{T}^{(1)}}$, is a nonterminating series.
However, as described earlier, in the present calculations we consider upto 
second order in $T$. The contributions from the higher order terms, based on 
previous studies with an iterative all order method\cite{mani-10}, is 
negligible. The contributions from different terms in 
Eq. (\ref{alpha_eqn}) are listed in Table. \ref{ne_result3}.
\begin{table}[h]
    \caption{Contribution to $\alpha $ of Ne from different terms of the 
             dressed dipole operator in the linearized PRCC theory}
    \label{ne_result3}
    \begin{center}
    \begin{tabular}{ld}
        \hline
        Contributions from  & \multicolumn{1}{c}{$\alpha$} \\  
        \hline
        $\{\contraction[0.5ex]{}{T}{_{1}^{(1)\dagger}}{D}
        \mathbf{T}_1^{(1)\dagger}\mathbf{D} \}$ + h.c.& 2.6610 \\
        $\{ \contraction[0.5ex]{}{\mathbf{T}}
        {_1{^{(1)\dagger}}}{D}\contraction[0.8ex]{}{\mathbf{T}}
	{{_1{^{(1)\dagger}}}D}{T}\mathbf{T}_1{^{(1)\dagger}}
	\mathbf{D}T_2^{(0)} \}$  + h.c. & -0.0478 \\
        $\{ \contraction[0.5ex]{}{\mathbf{T}}{_1{^{(1)\dagger}}}{D}
        \contraction[0.8ex]{}{\mathbf{T}}{{_1{^{(1)\dagger}}}D}{T}
        \mathbf{T}_1{^{(1)\dagger}}\mathbf{D}T_1^{(0)} \}$ + h.c. & 0.0644 \\
        $\{ \contraction[0.5ex]{}{\mathbf{T}}{_2{^{(1)\dagger}}}{D}
        \contraction[0.8ex]{}{\mathbf{T}}{{_2{^{(1)\dagger}}}D}{T}
        \mathbf{T}_2{^{(1)\dagger}}\mathbf{D}T_1^{(0)} \}$  + h.c. & -0.0062 \\
        $\{ \contraction[0.5ex]{}{\mathbf{T}}{_2{^{(1)\dagger}}}{D}
        \contraction[0.8ex]{}{\mathbf{T}}{{_2{^{(1)\dagger}}}D}{T}
        \mathbf{T}_2{^{(1)\dagger}}\mathbf{D}T_2^{(0)} \}$ + h.c. & 0.0961 \\ 
        Normalization & 1.0367 \\
        Total & 2.6695 \\
        \hline
    \end{tabular}
    \end{center}
\end{table}
  As evident from the table, the dominant contribution 
arises from $\{\contraction[0.5ex]{}{T}{_{1}^{(1)\dagger}}{D}\mathbf{T}_1^{(1)
\dagger}\mathbf{D} \}$ and its hermitian conjugate. This is not surprising as 
these terms subsume the DF contribution and core-polarization effects. The 
general trend is, for closed-shell atoms, the DF and core-polarization effects
are the leading order and next to leading order, respectively. Coming to the 
pair correlation effects, the leading contribution arise from 
$\{ \contraction[0.5ex]{}{\mathbf{T}}{_2{^{(1)\dagger}}}{D} 
\contraction[0.8ex]{}{\mathbf{T}}{{_2{^{(1)\dagger}}}D}{T}
\mathbf{T}_2{^{(1)\dagger}}\mathbf{D}T_2^{(0)} \}$ and its hermitian conjugate.
This is along the expected lines as the $T^{(0)}_2$ amplitude is larger, 
compared to $T_1^{(0)} $, on account of pair-correlations. The contributions
from the remaining terms are small and cancellations reduce the combined 
contribution even further.

  The next level of calculation is to consider all the terms in the
Eq. (\ref{prcc_eq1}) and (\ref{prcc_eq2}), which we refer to as the non-linear
PRCC.
\begin{table}[h]
   \caption{Contribution from all the terms in the nonlinear PRCC theory.}
   \label{ne_nprcc}
   \begin{center}
   \begin{tabular}{ld}
       \hline
       CC terms & \multicolumn{1}{c}{$\;\;\;\;\;\;\alpha$} \\
       \hline
       $\{\contraction[0.5ex]{}{T}{_{1}^{(1)\dagger}}{D}
       \mathbf{T}_1^{(1)\dagger}\mathbf{D} \}$ + h.c. & 2.7344 \\
       $\{ \contraction[0.5ex]{}{\mathbf{T}}
       {_1{^{(1)\dagger}}}{D}\contraction[0.8ex]{}{\mathbf{T}}
       {{_1{^{(1)\dagger}}}D}{T}\mathbf{T}_1{^{(1)\dagger}}
       \mathbf{D}T_2^{(0)} \}$ + h.c. & -0.0492 \\     
       $\{ \contraction[0.5ex]{}{\mathbf{T}}{_1{^{(1)\dagger}}}{D}
       \contraction[0.8ex]{}{\mathbf{T}}{{_1{^{(1)\dagger}}}D}{T}
       \mathbf{T}_1{^{(1)\dagger}}\mathbf{D}T_1^{(0)} \}$ + h.c. & 0.0670\\
       $\{ \contraction[0.5ex]{}{\mathbf{T}}{_2{^{(1)\dagger}}}{D}
       \contraction[0.8ex]{}{\mathbf{T}}{{_2{^{(1)\dagger}}}D}{T}
       \mathbf{T}_2{^{(1)\dagger}}\mathbf{D}T_1^{(0)} \}$ + h.c. & -0.0058 \\
       $\{ \contraction[0.5ex]{}{\mathbf{T}}{_2{^{(1)\dagger}}}{D}
       \contraction[0.8ex]{}{\mathbf{T}}{{_2{^{(1)\dagger}}}D}{T}
       \mathbf{T}_2{^{(1)\dagger}}\mathbf{D}T_2^{(0)} \}$ + h.c. & 0.0924 \\ 
       Normalization & 1.0367 \\
       Total & 2.7383 \\
       \hline
   \end{tabular}
   \end{center}
\end{table}
The term wise contributions are listed in Table. \ref{ne_nprcc} and the net
result of $2.7383$ is 2.6\% larger than the linearized PRCC result. 
 As evident from the table, most of the change is attributed to
$\{\contraction[0.5ex]{}{T}{_{1}^{(1)\dagger}}{D}
\mathbf{T}_1^{(1)\dagger}\mathbf{D} \}$  and hermitian conjugate. Contribution
from this term is 2.7\% larger in the nonlinear PRCC, which is comparable to
the change in the value of $\alpha$. This is one of the case where higher 
order calculations does not translate into improved accuracy. A similar 
situation, but in a different context, was observed in a detailed 
analysis of contributions from nonlinear terms in the CCSD and dressing
to calculate the magnetic dipole hyperfine constant of Li
\cite{derevianko-08}. Like in the work referred, the contributions from higher 
order cluster operators, triple and quadruple excitations, could be of 
different phase and bring $\alpha $ closer to experimental data. However, to 
examine or confirm this requires more detailed analysis and is beyond the 
scope of the present work.
\begin{table}[h]
   \caption{Two of the leading order terms in the nonlinear PRCC Theory. }
   \label{Ne_2_3}
   \begin{center}
   \begin{tabular}{ldd} \hline
       Contributions & 
       \multicolumn{1}{c}{$\{\contraction[0.5ex]{}{H}{_{\rm N}}{T}
       \contraction[0.8ex]{}{H}{_{\rm N}T^{(0)}}{T} 
       H_{\rm N}T_2^{(0)}\mathbf{T}_1^{(1)}\}$} & 
       \multicolumn{1}{c}{$\{\contraction[0.5ex]{}{H}{_{\rm N}}{T}
       \contraction[0.8ex]{}{H}{_{\rm N}T^{(0)}}{T} 
       \contraction[1.1ex]{}{H}{_{\rm N}T^{(0)}T^{(0)}}{T} 
       H_{\rm N}T_1^{(0)}T_1^{(0)}\mathbf{T}_2^{(1)}\}$} \\ \hline
       $\{\contraction{}{T}{_{1}^{(1)\dagger}}{D}
       \mathbf{T}_1^{(1)\dagger}\mathbf{D} \} + \text{h.c}$ & 
              2.7456 & 2.6628 \\
       $\{ \contraction[0.5ex]{}{\mathbf{T}}{_1{^{(1)\dagger}}}{D}
       \contraction[0.8ex]{}{\mathbf{T}}{{_1{^{(1)\dagger}}}D}{T}
       \mathbf{T}_1{^{(1)\dagger}}\mathbf{D}T_2^{(0)} \} + \text{h.c.}$ & 
              -0.0492 & -0.0478\\
       $\{ \contraction[0.5ex]{}{\mathbf{T}}{_1{^{(1)\dagger}}}{D}
       \contraction[0.8ex]{}{\mathbf{T}}{{_1{^{(1)\dagger}}}D}{T}
       \mathbf{T}_1{^{(1)\dagger}}\mathbf{D}T_1^{(0)} \} + \text{h.c.}$ & 
                0.0674 & 0.0642 \\
       $\{ \contraction[0.5ex]{}{\mathbf{T}}{_2{^{(1)\dagger}}}{D}
       \contraction[0.8ex]{}{\mathbf{T}}{{_2{^{(1)\dagger}}}D}{T}
       \mathbf{T}_2{^{(1)\dagger}}\mathbf{D}T_1^{(0)} \} + \text{h.c.}$ & 
               -0.0058 & -0.0058\\
       $\{ \contraction[0.5ex]{}{\mathbf{T}}{_2{^{(1)\dagger}}}{D}
       \contraction[0.8ex]{}{\mathbf{T}}{{_2{^{(1)\dagger}}}D}{T}
       \mathbf{T}_2{^{(1)\dagger}}\mathbf{D}T_2^{(0)} \} + \text{h.c.}$ & 
                0.0933 & 0.0922\\
       Normalization & 1.0367 & 1.0367 \\
       Total & 2.7503 & 2.6677 \\
       \hline
   \end{tabular}
   \end{center}
\end{table}

  Through a series of rigorous calculations, we examine the changes in 
$\alpha$, and associate it with a nonlinear term in Eq. (\ref{prcc_eq1}) and 
(\ref{prcc_eq2}). At the second order, there is an anomalously large 
contribution from 
$\{\contraction[0.5ex]{}{H}{_{\rm N}}{T}\contraction[0.8ex]{}{H}{_{\rm N}
T^{(0)}}{T} H_{\rm N}T_2^{(0)}\mathbf{T}_1^{(1)}\}$, it induces a changes of 
$0.0808$ a.u. to the net result of $\alpha$. This term accounts for the large
change of $\alpha$ in the nonlinear PRCC calculations. Compared to this term, 
the contribution from other terms at this order are marginal. The next largest 
contribution arises from 
$\{\contraction[0.5ex]{}{H}{_{\rm N}}{T} \contraction[0.8ex]{}{H}{_{\rm N}
T^{(0)}}{T} H_{\rm N}T_1^{(0)}\mathbf{T}_2^{(1)}\}$, it contributes
0.0086 a.u. The other contributions are  
0.0004 and 0.0034 a.u. from 
$\{\contraction[0.5ex]{}{H}{_{\rm N}}{T}\contraction[0.8ex]{}{H}
{_{\rm N}T^{(0)}}{T}H_{\rm N}T_1^{(0)} \mathbf{T}_1^{(1)}\}$ and
$\{\contraction[0.5ex]{}{H}{_{\rm N}}{T}\contraction[0.8ex]{}{H}{_{\rm N}
T^{(0)}}{T} H_{\rm N}T_2^{(0)}\mathbf{T}_2^{(1)}\}$, respectively.

 At the third order $\{\contraction[0.5ex]{}{H}{_{\rm N}}{T}
\contraction[0.8ex]{}{H}{_{\rm N}T^{(0)}}{T}
\contraction[1.1ex]{}{H}{_{\rm N}T^{(0)}T^{(0)}}{T} 
H_{\rm N}T_1^{(0)}T_1^{(0)}\mathbf{T}_1^{(1)}\}$ 
and 
$\{\contraction[0.5ex]{}{H}{_{\rm N}}{T}
\contraction[0.8ex]{}{H}{_{\rm N}T^{(0)}}{T} 
\contraction[1.1ex]{}{H}{_{\rm N}T^{(0)}T^{(0)}}{T} 
H_{\rm N}T_1^{(0)}T_2^{(0)}\mathbf{T}_2^{(1)}\}$
contribute equally, 0.0077 a.u. each. The contribution from  the last term
at this order,
$\{\contraction[0.5ex]{}{H}{_{\rm N}}{T}
\contraction[0.8ex]{}{H}{_{\rm N}T^{(0)}}{T} 
\contraction[1.1ex]{}{H}{_{\rm N}T^{(0)}T^{(0)}}{T} 
H_{\rm N}T_1^{(0)}T_1^{(0)}\mathbf{T}_2^{(1)}\}$, is $-0.0018$ a.u.
To illustrate the relative changes arising from the third order terms, we
list the contributions from the leading order terms in the second order
and third order in Table. \ref{Ne_2_3}. It is evident from the table 
that the difference between the second and third order
contributions arises from the 
$\{\contraction[0.5ex]{}{T}{_{1}^{(1)\dagger}}{D}
\mathbf{T}_1^{(1)\dagger}\mathbf{D} \}$ and its hermitian conjugate.

  At the fourth order there is only one term 
$\{\contraction[0.5ex]{}{H}{_{\rm N}}{T}
\contraction[0.8ex]{}{H}{_{\rm N}T^{(0)}}{T} 
\contraction[1.1ex]{}{H}{_{\rm N}T^{(0)}T^{(0)}}{T} 
\contraction[1.4ex]{}{H}{_{\rm N}T^{(0)}T^{(0)}T^{(0)}}{T} 
H_{\rm N}T_1^{(0)}T_1^{(0)}T_1^{(0)}\mathbf{T}_1^{(1)}\} $ and contributes
0.0077 a.u. This detailed study implies that the higher order terms in
the PRCC equations, third and fourth order, have negligible effect on the
electric dipole polarizability. Since the effect of the higher terms 
are tightly coupled to the electron correlation effects, a similar trend may 
occur in other properties as well.
\begin{table}[h]
  \caption{The contribution to $\alpha$ from the fourth order term in nonlinear
           PRCC theory.}
  \label{Ne}
  \begin{center}
  \begin{tabular}{lc}
     \hline
     Contributions From & 
     $\{\contraction{}{H}{_{\rm N}}{T}
     \contraction[1.5ex]{}{H}{_{\rm N}T^{(0)}}{T} 
     \contraction[2.0ex]{}{H}{_{\rm N}T^{(0)}T^{(0)}}{T} 
     \contraction[2.5ex]{}{H}{_{\rm N}T^{(0)}T^{(0)}T^{(0)}}{T} 
     H_{\rm N}T_1^{(0)}T_1^{(0)}T_1^{(0)}\mathbf{T}_1^{(1)}\} $ \\
     \hline
     $\{\contraction[0.5ex]{}{T}{_{1}^{(1)\dagger}}{D}
     \mathbf{T}_1^{(1)\dagger}\mathbf{D} \} + \text{h.c.}$ & 2.6688 \\
     $\{ \contraction[0.5ex]{}{\mathbf{T}}{_1{^{(1)\dagger}}}{D}
     \contraction[0.8ex]{}{\mathbf{T}}{{_1{^{(1)\dagger}}}D}{T} 
     \mathbf{T}_1{^{(1)\dagger}}\mathbf{D}T_2^{(0)}\} + \text{h.c.} $ 
                 & -0.0478 \\ 
     $\{ \contraction[0.5ex]{}{\mathbf{T}}{_1{^{(1)\dagger}}}{D}
     \contraction[0.8ex]{}{\mathbf{T}}{{_1{^{(1)\dagger}}}D}{T}
     \mathbf{T}_1{^{(1)\dagger}}\mathbf{D}T_1^{(0)} \} + \text{h.c}$ 
                 & 0.0645 \\
     $\{ \contraction[0.5ex]{}{\mathbf{T}}{_2{^{(1)\dagger}}}{D}
     \contraction[0.8ex]{}{\mathbf{T}}{{_2{^{(1)\dagger}}}D}{T}
     \mathbf{T}_2{^{(1)\dagger}}\mathbf{D}T_1^{(0)} \} + \text{h.c}$ 
                 & -0.0062 \\
     $\{ \contraction[0.5ex]{}{\mathbf{T}}{_2{^{(1)\dagger}}}{D}
     \contraction[0.8ex]{}{\mathbf{T}}{{_2{^{(1)\dagger}}}D}{T}
     \mathbf{T}_2{^{(1)\dagger}}\mathbf{D}T_2^{(0)} \} + \text{h.c}$ 
                 & 0.0962 \\ 
     Normalization & 1.0367 \\
     Total & 2.6772 \\
     \hline
  \end{tabular}
  \end{center}
\end{table}

To estimate the uncertainty in our calculations, we have identified two
sources in the calculations using PRCC with CCSD approximation.
First type of error is associated with the orbital basis set truncation
and the termination of iteration while solving the cluster amplitudes. 
Based on the basis set convergence, as described earlier, the uncertainty 
from the basis set truncation is negligible. Similarly, the uncertainty from 
the termination of cluster amplitude calculation is negligible as we set
$10^{-6}$ as the convergence criterion.  The second type of error arises
from the truncation of the CC theory at double excitations and the 
truncation of $e^{{\mathbf{T}^{(1)}}^\dagger}De^{T^{(0)}} 
+ e^{{T^{(0)}}^\dagger}De^{\mathbf{T}^{(1)}}$. Based on other detailed 
studies on the contributions from the triples and quadruple excitations
could be in the range of $\approx$-2.6\%. So that it balances the larger
error arising from the inclusion of the nonlinear terms in the PRCC theory.
Again, based on other studies with iterative method \cite{mani-10} to 
incorporate higher order terms in the properties calculations
with CC theory, the contributions from the third or higher order in 
$e^{{\mathbf{T}^{(1)}}^\dagger}De^{T^{(0)}} 
+ e^{{T^{(0)}}^\dagger}De^{\mathbf{T}^{(1)}}$ is negligibly small. 
The contribution from Breit and QED corrections could be another source of 
error. However, as $Z\alpha\ll 1$, where $\alpha$ is the fine structure 
constant for Ne the uncertainty from excluding Breit and QED correction could 
easily be sub 0.01\%. This is consistent with the estimates of the contribution
from the Breit interaction to correlation energy \cite{ishikawa-94}. Combining 
all the sources of error, the uncertainty for the calculations with nonlinear 
PRCC is $\approx$ 2.6\%. But due to possible fortuitous cancellations, the 
uncertainty with linearized PRCC calculations is sub 0.1\%.

\section{conclusion}

  The PRCC theory provides a consistent approach within CC theory to 
calculate atomic properties. Although, in this paper we have demonstrated
the use of PRCC to calculate $\alpha$ of Ne. The
method is applicable to any atomic property. In PRCC theory, the number of 
cluster amplitudes is larger than the RCC theory but the PRCC equations are 
linear. The main feature of the method is accounting of all possible 
intermediate states within the basis set chosen for the calculations. With 
converged basis sets, the intermediate states incorporated in the properties 
calculations is practically complete.

  From the detailed calculations presented in this work, it can be concluded 
that inclusion of the nonlinear terms does not improve the accuracy 
of the properties calculated with PRCC theory. However, The linearized PRCC 
provides results which are in excellent agreement with the experimental data. 
The lower accuracy with the nonlinear PRCC may be attributed to large
cancellations between contributions from different nonlinear terms. As to
be expected, the leading order term is
$\{\contraction[0.5ex]{}{T}{_{1}^{(1)\dagger}}{D}\mathbf{T}_1^{(1)
\dagger}\mathbf{D} \}$. It accounts for, in the linearized PRCC calculations,
96\% of the total value.

\begin{acknowledgements}
We thank S. Gautam, Arko Roy and Kuldeep Suthar for useful discussions. The 
results presented in the paper are based on the computations using the 3TFLOP 
HPC Cluster at Physical Research Laboratory, Ahmedabad. 
\end{acknowledgements}

\bibliography{dipole_polarizability}{}
\bibliographystyle{apsrev4-1}

\end{document}